# Towards Usability Guidelines for the Design of Effective Arabic Websites

## Design Practices and Lessons Focusing on Font and Image usage

Abdallah Namoun[1]
Department of Information Systems
Faculty of Computer and Information Systems
Madinah, Saudi Arabia

Ahmad B. Alkhodre[2]
Department of Information Technology
Faculty of Computer and Information Systems
Madinah, Saudi Arabia

*Abstract*—The Arabic websites constitute 1% of the web content with more than 225 million viewers and 41% Internet penetration. However, there is a lack of design guidelines related to the selection and use of appropriate font type and size and images in Arabic websites. Both text and images are vital multimedia components of websites and thereby were selected for investigation in this study. The herein paper performed an in-depth inspection of font and image design practices within 73 most visited Arabic websites in Saudi Arabia according to Alexa Internet ranking in the first quarter of 2019. Our exhaustive analysis showed discrepancies between the international design recommendations and the actual design of Arabic websites. There was a considerable variation and inconsistency in using font types and sizes between and within the Arabic websites. Arabic Droid Kufi was used mostly for styling page titles and navigation menus, whilst Tahoma was used for styling paragraphs. The font size of the Arabic text ranged from 12 to 16 pixels, which may lead to poor readability. Images were used heavily in the Arabic websites causing prolonged site loading times. Moreover, the images strongly reflected the dimensions of the Saudi culture, especially collectivism and masculinity. Current Arabic web design practices are compared against the findings from past studies about international designs and lessons aiming at ameliorating the Arabic web design are inferred.

## I. INTRODUCTION

There is growing evidence in the literature that web design and content significantly impact the perceived usability and acceptability of websites [1]. For instance, reference [2] stipulated that the visual appeal of websites affects users' perception of the website and that judgment is usually formed within the first 50 milliseconds of exposure. Actually, 70% of the web content is represented in the form of text or images which are considered the building blocks of any website [3]. Limited research efforts attempted to address the question of 'what is the optimal way to represent textual content and graphics within Arabic websites.

Both text typography [4] and images [5] are believed to enhance content understanding and predict user perception, acceptance and engagement. However, the factors that affect the acceptability of Arabic websites, which constitute 1% of the web, remain unexplored. Font type and size influence reading speed and overall page appearance, whilst images and graphics impact the visual appeal of the site.

In this study, we are faced with two web design questions and research challenges. What font types and font sizes are used by major Arabic websites? Determining these typographical characteristics will assist us to understand the level of conformance to international design recommendations and best practices for web design. What types of images are used within the Arabic websites given the Saudi culture, values, and religious identity? The proverb says "a picture is worth a thousand words"; however, it is important to ensure that the images used in web design conform to the context and society where they are being introduced. Our inspection of images' characteristics will assist us in identifying the common practices within the Saudi environment.

Studies investigating the effects of fonts and images on user judgment and acceptance of Arabic websites are limited in the literature. To this end, it is crucial to first identify the common fonts and images being used by Arabic websites. This paper focuses on exploring the current practices adopted by designers for styling Arabic text with regard to the font type and size, as well as the genre of images incorporated within the Arabic websites. Such knowledge lays the foundations for further web design research.

The paper is divided into seven main sections. Section two sets the foundations of usability and web design and reviews past studies focusing on web typography and images. Section three describes the research methodology we adopted to answer the research questions. Section four presents the results of our in-depth inspection of current font and image practices. Section five discusses the findings and predicts potential implications. Section six and seven provide an outlook towards potential ways to enhance the Arabic web design.

## II. PREVIOUS WORKS

### A. Perceived Usability

Online companies that provide a rich user experience by accounting for website usability are more likely to succeed than those which do not [1], [6]. In other words, unusable websites drive customers away and cause them to look for alternative websites [7]. Perceived web usability is a quality factor that refers to the empowerment of users to achieve their goals and needs by exploiting the design features of the website [6]. Perceived usability is believed to contribute towards accepting and using a particular technology; for instance, [8] demonstrated that perceived usability impacts the acceptance of educational technologies by teachers.

Perception of website usability has been linked to numerous factors including attractiveness, controllability, efficiency, learnability, and helpfulness [33]. Moreover, users 'demographics are found to influence overall website perception. Further research reported a consolidated model of usability metrics and measurements [9]. This model encompassed 10 factors which are, in turn, divided into 126 metrics. The proposed quality in use integrated measurement model included efficiency, effectiveness, productivity, satisfaction, learnability, safety, trustfulness, accessibility, universality, and usefulness [9]. The readability of the text is indicated as one of the key metrics contributing to perceived usability.

### B. Web Design and Content

So what actually affects web design quality and overall usability perception of customers? Research has indicated that various factors impact how users perceive and interact with online websites. Indeed, both web content and web design have been considered as critical factors leading to the success of websites and repeated customer visits [3], [1]. Author in [2] defined content as the information or services available on a particular website whilst design refers to the way this content is presented to the users. For example, [1] proposed a web usability assessment model that incorporates multiple factors such as navigation, content, and layout. Evidence in the literature confirms that when textual content is readable easily, the website receives higher usability by its viewers [10].

Authors in [10] and [11] carried out three website studies to explore the relationship between the perceived usability of websites, web design and performance. Findings showed that website success and satisfaction are highly correlated with the available content, navigational structure, interactivity, feedback, and site loading speed. Author in [7] presented an automated tool that checks the conformance of websites to design guidelines and predicts an overall website rating. The usability checks assess the quality of the web design by measuring aspects related to the text, links, graphics, as well as their formatting. However, the model neither investigated Arabic fonts nor the types of images used.

Author in [12] found that web design is a critical success factor for positive user experience and interaction on websites. Effective web design contributes towards user acceptance and empowers users to find relevant information easily in e-commerce sites. Moreover, visual appearance is found to significantly influence users' perceptions of business to consumer websites [13].

Author in [14] described a detailed framework to assess the usefulness of web content. A total of 18 content benchmarking criteria (e. g. text quality, scope, accuracy, uniqueness) were collated from existing works and expert reviews. In respect to screen appearance, the authors suggested using different readable font sizes. Effective and easy navigational structures within a site are favored by website users than complex navigations [14].

### C. Font Type and Size

Font type and font size influence the readability and legibility of text. Recent research, for instance [47], demonstrated that difficult to read fonts encourage the willingness to pay for the adventure tours. A strong link between visual attributes of travel text and intention to travel was established. In addition, [49] showed that users' judgment of memory retention will be higher as a result of increasing the font size. In smartphones, the big text was liked more than smaller text when styling brand names [50].

Author in [1] advocated using adequate font size for displaying information on websites. However, determining the appropriate font type and size for text on websites is not an easy task. In an eye-tracking experiment, [15] demonstrated through eye gaze data that the speed of online reading deteriorates in the presence of small font sizes. Only Helvetica and Georgia fonts were compared at varied sizes 10, 12, and 14 points. Surprisingly, no significant differences were detected between these sizes and between the serif and non-serif fonts in respect to reading speed.

In contrast, Serif fonts performed better than Sans Serif with respect to online readability [4] and information recall [16] whilst Sans Serif fonts were preferred over Serif fonts [4]. Font type Verdana at size 14 points ranked first amongst online readers and induced the lowest mental load.

Author in [10] carried out an experiment to compare eight famous fonts, namely Century Schoolbook, Courier New, Georgia, Times New Roman, Arial, Comic Sans MS, Tahoma, and Verdana at size 10, 12, and 14 points respectively. Although these varied sizes did not result in different reading efficiency, Times New Roman and Arial were read significantly faster than the other fonts. Moreover, font size 10 was read significantly slower than font size 12 and 14. In respect to legibility, Arial and Courier were perceived more legible than the remaining fonts. Size, however, did not seem to improve font legibility. In respect to attractiveness, Georgia was rated as the most attractive font amongst the candidate fonts. The overall ranking of fonts showed Verdana as the most preferred font, whereas Times New Roman was ranked as the least preferred font type.

In a relevant study, students aged between 10 and 12 years participated in a screen legibility experiment of Arabic characters [4]. Only two font types were investigated, mainly Arabic Traditional and Simplified Arabic in different sizes 10, 14, 16 and 18 points respectively. Results indicated that font size 16 and 18 were more readable than font size 10 and 14 by 9-12 years aged students. However, font size 10 triggered the highest number of errors. Nonetheless, [4] claim that font type did not influence the readability of the Arabic language.

For the English language, font size 10 and 12 points were found to be equally readable [10]. However, for the Arabic characters [17] suggested that text is more readable at font size 14 points. Moreover, [17] claimed that font types influence reading speed, which contradicts the findings of [4]. In fact, author in [4] recommends using at least font size 18 point and over for efficient online reading of the Arabic text.

More recent research investigated how four fonts (i.e. Georgia, Verdana, Times New Roman, and Arial) impact online readability [18]. Verdana and Georgia were favored for computer displays while Times New Roman and Arial were favored for printed media. Reference [19] confirmed Verdana as a more suitable font type for showing English text on computerized screens.

### D. Images

Author in [14] emphasized the need to use images and graphical elements to convey messages and improve the overall usability of websites. Moreover, the use of graphics enhances the aesthetics of websites and leads to user satisfaction [18]. Visual appeal was deemed as important as content for visual arts websites [3]. Graphics are known to improve the overall perceived attractiveness of websites and design considerations must be catered for different cultures [20].

Author in [21] compared three different e-commerce web designs, a website containing no human images, a website containing human images without faces, and a website containing images with human faces to explore how the use of images influence visual appeal of websites. The websites utilizing images with human faces were perceived as more appealing and trustworthy. Reference [22] emphasized the need to take cultural differences into consideration when embedding graphical components within e-commerce websites.

### E. Design Guidelines

Author in [22] recommended the design of culturally sensitive websites, use of correct translation, employment of consistent colors, and minimization of animations. HHS design guidelines advised using at least font size 12 points and sticking to familiar fonts [23], [24]. For image usage, HHS design guidelines proposed to add images to the website and use images of people, but reduce their size so as not to slow down page loading. Reference [48] suggested the consideration of site loading time as a key usability metric as part of an evaluation framework of dual language websites.

### F. The Research Gap

Research efforts tackling the usability of Arabic websites are limited. Some studies examined the performance of Arabic websites by inspecting the internal attributes of the HTML pages. For example, [25] compared three Arabic educational websites and reported technical issues related to the site loading speed as a direct consequence of the excessive number of HTML objects and images and their large size. Similarly, [26] inspected the number of HTML files, objects, images, CSS files, broken links and size of images on 9 websites using two specialized online tools. Other aspects reported included pa`ge loading time, HTML errors and browser compatibility.

Until today and despite the prevalence of the Arabic language on the web, there are still no systematic studies that examine the use of fonts and images on Arabic websites. This is the first research effort to review and analyze a large number of websites to identify the real practices pertaining to the usage of fonts and images on Arabic websites.

### III. RESEARCH METHODOLOGY

Evaluation methods of websites' usability vary greatly in regard to the procedures, complexity, and accuracy [27]. Usability inspection is the second most used web usability evaluation method after user testing [27], [9]. In simple terms, usability inspection refers to the methods that are employed by an evaluator to inspect usability issues or features of a specific user interface [9]. In our case, we concentrate on two main design elements; firstly we inspect the quality of the typography of 73 Arabic websites and the genre of images used within these websites and secondly we verify the agreement of these design elements (i. e. typography and images) with the proposed international design guidelines. The steps of the evaluation we performed are depicted in Fig. 1.

This form of inspection is called 'perspective-based inspection' where the evaluator concentrates on particular design elements [28]. It is worthwhile to note that inspection differs from user testing for it does not involve actual users in the evaluation process. This makes it more favorable especially when the number of websites to test is large as is our case. Moreover, conducting usability inspection is still considered useful and more affordable than other evaluation techniques [29].

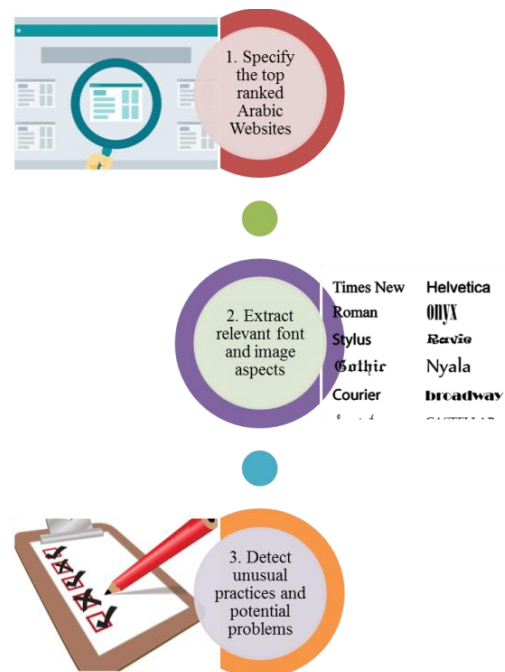

Fig. 1. Research Methodology.

## IV. FONT AND IMAGE INSPECTION STUDIES

### A. The Morphology of the Arabic Language

Before delving into the details of the studies conducted in this research, a brief introduction to the Arabic language is given. Arabic is one of the richest Semitic languages with more than 12 million unique words and 300 million speakers across the globe. Arabic is written from right to left. In essence, words are created from a set of 28 distinct letters, with 15 letters using dots, either above or below the letter (as depicted in Fig. 2). The Arabic letters use diacritics which create different meanings according to the context of use.

Table I demonstrates how applying well-known font types changes the look and feel and overall readability of the Arabic text. The selected fonts to showcase are the most used fonts as per our analysis below. This includes Times New Roman, Arial, Tahoma, Helvetica, Cairo, and Droid Arabic Kufi. All font types are applied to the same sentence and are sized to 8 points (~ 11 pixels).

### B. Selection of the Arabic Websites

A total of 73 differing Arabic websites were selected and evaluated on several metrics with a focus on font and image characteristics as detailed below. The selected online websites for investigation were the most visited websites by users living in the Kingdom of Saudi Arabia according to the latest ranking (i.e. February 2019) of Alexa Internet [30]. Technically, Alexa ranks websites locally or globally using multiple Internet browsing and web traffic criteria, including the number of monthly unique visitors to the site, number of site views for the last three months, and average time spent viewing the website. The websites we decided to investigate were the most visited Arabic websites and included 52 local websites (71.23%) and 21 international websites (28.77%). This enabled us to learn about the font and image related design practices of both local (e. g. Souq, Almubasher … etc.) and international sites (e. g. Google, YouTube, Microsoft … etc.). A selected sample of the tested Arabic websites is shown below in Fig. 3, Fig. 4, and Fig. 5.

Fig. 6 depicts the distribution of 73 top ranked Arabic websites according to their genre. Notably, Government websites were the most commonly visited websites among the Saudi population, with 14 websites. This was followed by news (13 websites), corporate (11), education (9), e-commerce (7), entertainment (5) and social media (4) websites. The remaining sites (12 websites) spread across diverse categories, such as charity, search engine, travel, and tourism, etc.

TABLE I. THE EFFECTS OF FONT TYPE ON THE READABILITY OF ARABIC TEXT

| Font Type | Arabic Text (All sizes are set to 8 points, ~ 11 pixels) |
|---|---|
| Times New Roman | أثر نوع الخط على سهولة قراءة النص |
| Arial | أثر نوع الخط على سهولة قراءة النص |
| Tahoma | أثر نوع الخط على سهولة قراءة النص |
| Helvetica | أثر نوع الخط على سهولة قراءة النص |
| Cairo | أثر نوع الخط على سهولة قراءة النص |
| Droid Arabic Kufi | أثر نوع الخط على سهولة قراءة النص |

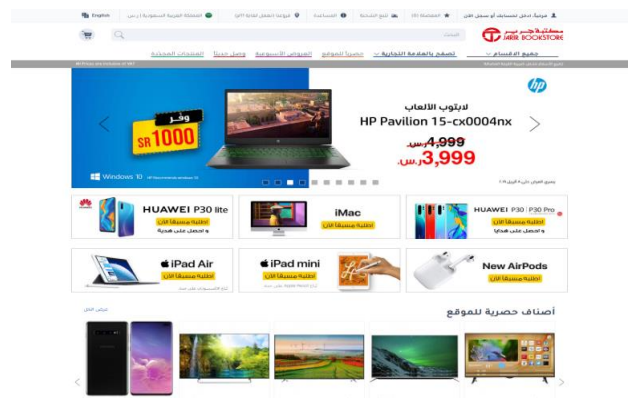

Fig. 2. The Arabic Characters and their Pronunciation, Adopted from [34].

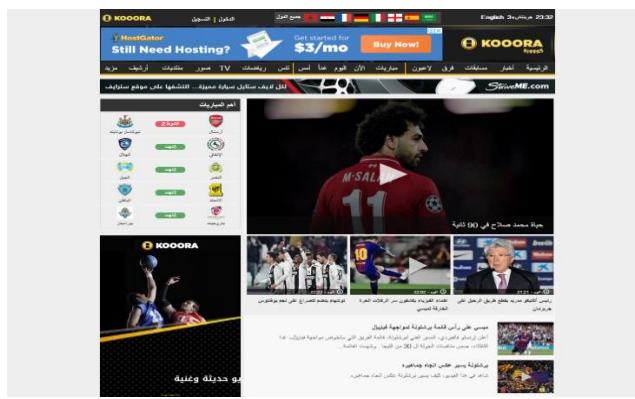

Fig. 3. Jarir Bookstore, An E-commerce Website.

Fig. 4. Kooraa, A News Website.

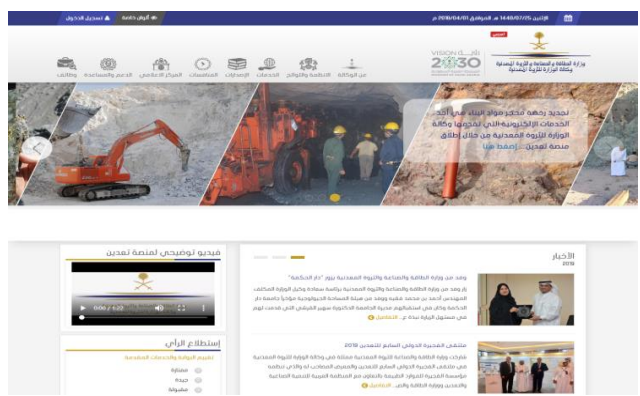

Fig. 5. Dmmr, A Government Website.

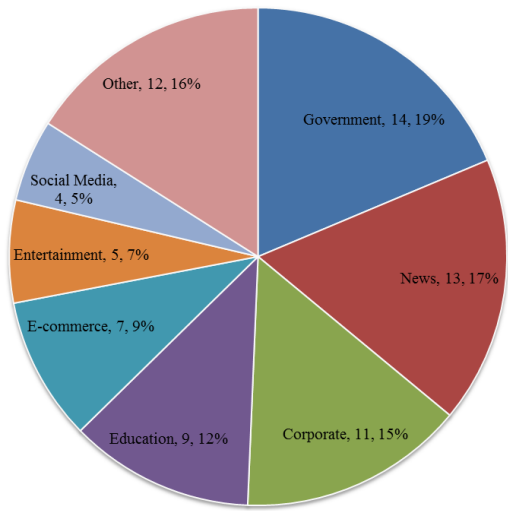

Fig. 6. The Categories of Arabic Websites Inspected and their Frequency
(Labels: Site Category, Frequency, Percentage).

## V. RESULTS

Our technical inspection focused on analyzing a set of characteristics that are related to font and image usage. Below we delve into the current practices of 73 top viewed Arabic websites.

### A. Practices Pertaining to Font Type and Font Size

Chrome extension Font Face Ninja [31] was installed to detect and identify the fonts utilized to format the sample Arabic websites, with a particular emphasis on font aspects such as type and size (reported in pixels). In the inspection, we explored font styling pertaining to three key text elements namely, titles (or headings), navigation menus, and paragraphs.

Formatting text with a particular font size has a direct impact on the overall readability and legibility of text on computer screens. Our next analysis explored the various fonts and sizes applied to format the Arabic text. It is worth noting that approximately 40 websites (~55% of the websites) used different font types within their pages raising concerns about internal consistency.

The second observation we noticed is the large variability in the font's usage across the Arabic websites. In total, 39 different font types were utilized to style the titles of the 73 Arabic websites. The most frequent font type used for formatting titles was Droid Arabic Kufi (7 occurrences). When styling the navigation menus, 38 different font types were used, with Droid Arabic Kufi being the most frequent font. However, 35 different fonts were used to style the paragraphs of 73 Arabic websites, with Tahoma appearing more than the remaining fonts.

When font types were ordered according to their occurrence, six main types emerged as clear winners specifically; Droid Arabic Kufi, Tahoma, Arial, Cairo, Helvetica and Times New Roman (see Fig. 7). In respect to page titles and navigation menus, Droid Arabic Kufi, Tahoma and Helvetica were used more frequently. However, for styling paragraphs, Tahoma, Helvetica, Arial and Times New Roman were more dominant (see Fig. 7).

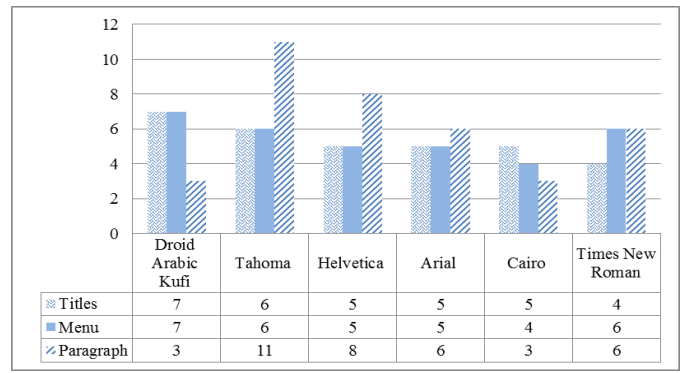

| | Droid Arabic Kufi | Tahoma | Helvetica | Arial | Cairo | Times New Roman |
|---|---|---|---|---|---|---|
| Titles | 7 | 6 | 5 | 5 | 5 | 4 |
| Menu | 7 | 6 | 5 | 5 | 4 | 6 |
| Paragraph | 3 | 11 | 8 | 6 | 3 | 6 |

Fig. 7. The Most Frequently used Font Types in Arabic Websites.

With regard to the font size used for the titles, size 28 pixels (8 times), 24 pixels (7 times) and 30 pixels (6 times) emerged as the most recurring sizes. Due to the variability of font sizes used for styling the Arabic titles, we created a histogram with the bins ranging from 10 to 14 pixels, 15 to 19 pixels, 20 to 24 pixels, 25 to 29 pixels, 30 to 34 pixels, 35 to 40 pixels, and more than 40 pixels (see Fig. 8). The below chart shows that the majority of Arabic websites styled their titles using sizes ranging from 15 to 40 pixels. Font sizes less than 15 pixels or more than 40 pixels were scarce.

Less font size variability was identified in relation to the formatting of menus of Arabic websites. In total, only 10 font sizes were utilized in the 73 websites, with font size ranging from 11 pixels to 22 pixels. The most recurring font size for formatting navigation menus was 14 pixels (19 times ~ 26%), followed by 13 pixels font size (13 times ~ 18%) (see Fig. 9).

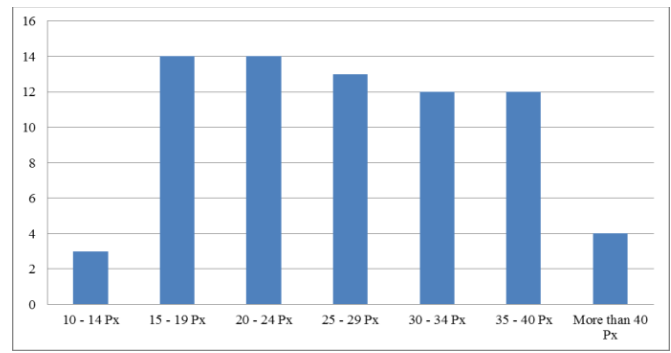

Fig. 8. The Frequency of Titles' Font Sizes (in Pixels) used in Arabic Websites.

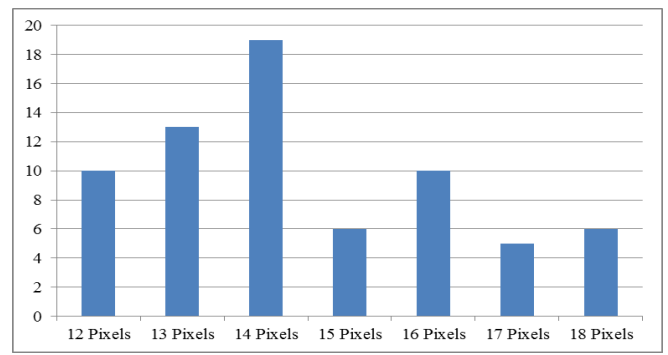

Fig. 9. The Frequency of Menu's Font Sizes (in Pixels) used in Arabic Websites.

When it came to formatting Arabic paragraphs, font size 14 pixels (~26%), 16 pixels (~19%), and 12 pixels (~18%) were more commonly used than the other sizes, as depicted in Fig. 10. Once again, 10 different font sizes, ranging from 12 to 26 pixels, were used to style the paragraphs of 73 websites. Only three websites used font sizes larger than 18 pixels to format the paragraphs.

### B. Practices Pertaining to Images and Graphics

In respect to the use of images, the inspection looked into the content of the websites' images with regard to four major aspects, the use of images containing human faces, the gender portrayed in the images, the use of individuals or groups, and the conservatism or openness of the image content with respect to the Saudi values and Islamic religion.

Moreover, a technical analysis was undertaken to check the number of images incorporated in each site and their size since these aspects have a direct impact on the loading speed of the web pages. We have used website speed test [32] to gather data about several aspects of the images and learn how these images could be optimized to ameliorate the overall site performance. The results showed that all 73 Arabic websites included images. A total of 3486 images (mean= 47.75 per page) were reported and analyzed by a website speed test. These analyzed images were fetched mainly from the home pages.

The average number of images varied greatly between the site categories. News, entertainment, government and e-commerce websites used approximately 50 images or more on their home pages respectively (Fig. 11). This is an anticipated result. The average size of the images ranged between 135KB and 2900KB. Most of the Arabic websites used images greater than 1500KB in size (Fig. 12). Social media websites produced smaller sizes as the analysis focused on the home pages only.

Overall, 66 (90%) Arabic websites embedded images that contained humans and only 7 websites did not include images with humans in them (see Table II). Interestingly, the vast majority of these websites (52 websites) used images that are conservative in nature respecting the Islamic identity and cultural values of the Saudi society. Nonetheless, 14 websites used non-conservative images, for example, pictures of women wearing short clothes. These images are considered unacceptable Islamically and against the values of the Saudi culture. All non-human images were conservative in nature and did not portray things against the teachings of Islam or Saudi society.

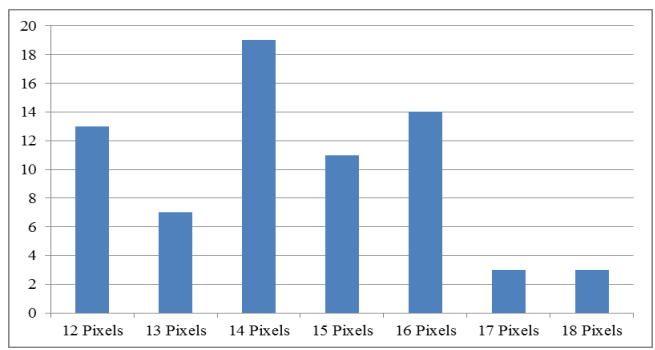

Fig. 10. The Frequency of Paragraph's Font Sizes (in Pixels) used in Arabic Websites.

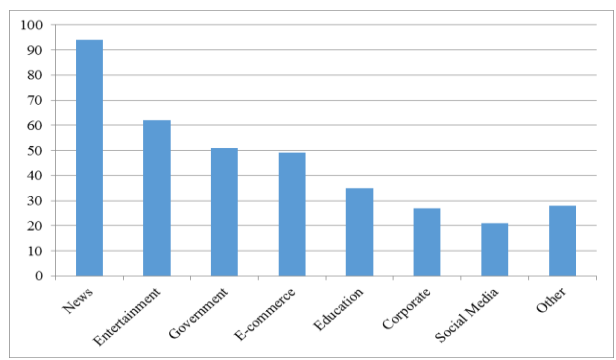

Fig. 11. Average Number of Images on Home Pages of the Arabic Websites.

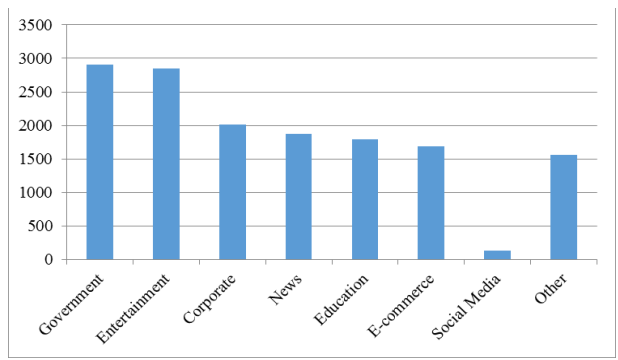

Fig. 12. Average Size of Images (in Kilobytes) on Home Pages of the Arabic Websites.

Moreover, 45 websites (out of 66) used images of men and women; this is approximately 68% of the websites including images containing humans (see Fig. 13). However, 21 websites (approximately 32%) used images of men only which shows that the masculinity in Saudi society still prevails. No website used images containing women alone.

TABLE II. THE TYPE OF IMAGES USED IN THE ARABIC WEBSITES

| Type | Conservative | Open | Total |
|---|---|---|---|
| Images containing humans | 52 | 14 | 66 (90.41%) |
| Images without humans | 7 | 0 | 7 (9.58%) |
| | 59 (80.82%) | 14 (19.18%) | |

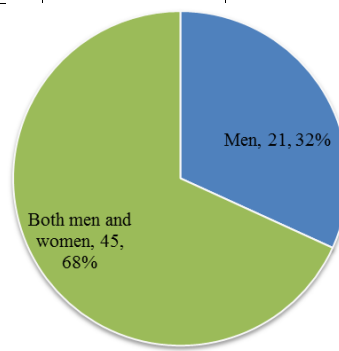

Fig. 13. The Frequency of Arabic Websites using Images Containing Men, Women or Both (Label: Gender, Frequency, Percentage).

48 Arabic websites (~72%) used a combination of images containing both individuals and groups of people (Fig. 14). This reflects the collective nature of Saudi society. In contrast, only 7 websites included pictures of individuals.

A total of 47 websites (~64%) incorporated some form of interactive animations ranging from rotating images, loading animations, to CCS animations (see Fig. 15). However, only 20 Arabic websites (~27%) incorporated videos.

## C. The Performance of the Arabic Websites

We undertook a general site analysis to check the performance of the Arabic websites. To this end, GTmetix[1] was used to collect useful data about the speed scores of the websites, the site loading time, and the total page size. The data were collected for all 73 Arabic websites and averaged per category as shown in Fig. 16 and Fig. 17. Site performance scores are rated as a relative percentage (100%) to the performance of other websites analyzed by GTmetrix in the last 30 days. Only social media and entertainment websites scored above than 70% of other sites. However, government, education, and news websites scored less than 50% of the other analyzed sites (see Fig. 16).

In respect to the site loading time, GTmetrix extracts metrics that capture the average time spent, in seconds, until the website ceases the data transfer process over the network and the onload events are launched on the designated website. The shorter this time is the better it is for the end user. Reduced loading time is also an indication of good implementation practices (e.g. fewer HTTP requests, compressed files…etc.). Author in [25] suggested that the site should take less than 3 seconds to load fully otherwise users might leave the site.

Our analysis showed that all categories took, on average, over 5 seconds to load except the social media sites (approximately 3 seconds). Social media loaded fast for they contain a low number of images on their homes pages prior to signing in. Interestingly, government, education, and news websites showed the worst performance taking 10 seconds or longer to fully load (as depicted in Fig. 17). The government websites contained large images as shown earlier in Fig. 12.

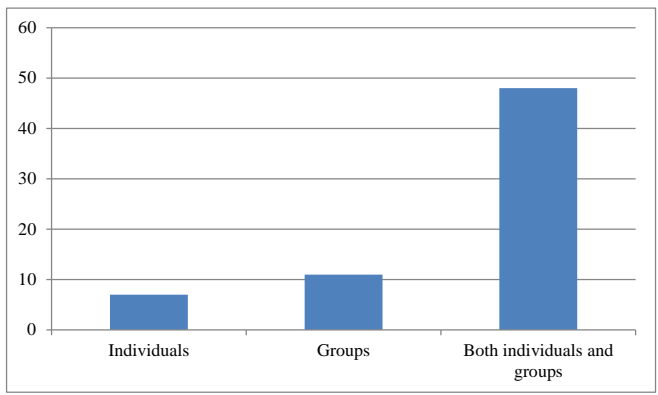

Fig. 14. The Frequency of Arabic Websites using Images Containing Individuals, Groups of People or Both.

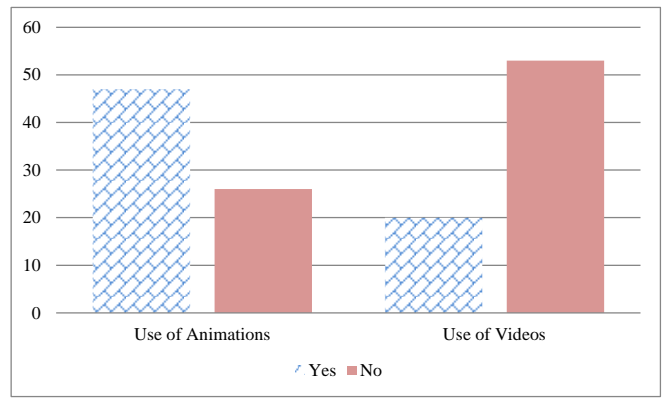

Fig. 15. The Frequency of Arabic Websites using Animations and Videos.

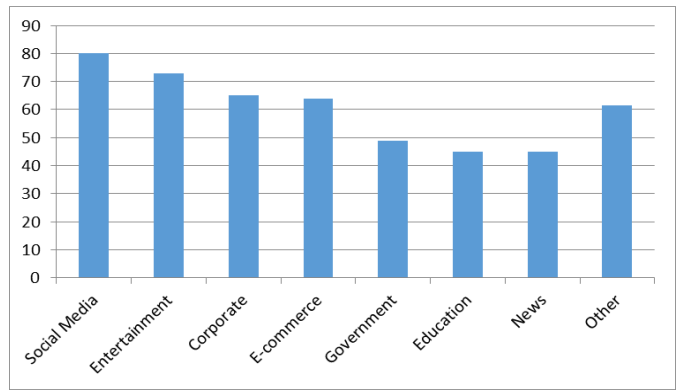

Fig. 16. Average Page Speed Score (as Percentage) per Site Category (High Scores are Better).

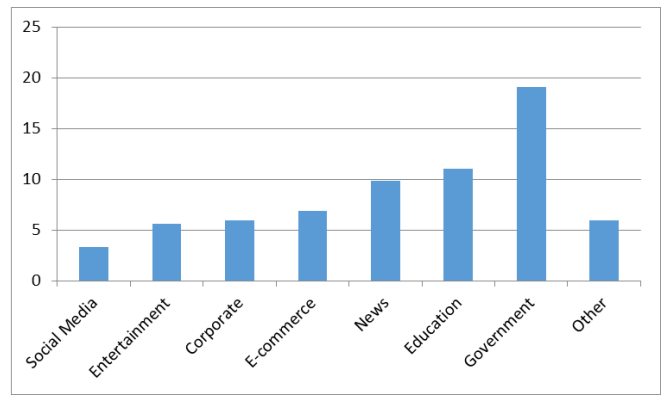

Fig. 17. Average Fully Loaded Time (in Seconds) per Site Category (More is Worse).

## VI. Discussion

Our comprehensive inspection of current practices concerning text formatting and image usage within the top visited Arabic websites in Saudi Arabia revealed various interesting findings. The usability inspection was performed to gather insights about the font types and sizes used to style Arabic text, especially that they could directly influence the legibility of text on websites.

[1] https://gtmetrix.com/

Let us review and attempt to answer the research questions posited in the introduction. What font types and font sizes are used by major Arabic websites? On the positive side, the selection of the Arabic websites included a large pool of websites, which were selected objectively by relying on Alexa Internet latest ranking. In effect, the inspection covered a total of 73 top viewed websites (52 local and 21 international) in Saudi Arabia. The websites' ranking is based on a set of metrics, primarily the average viewing time for the last three months.

Typically, a font has several characteristics including type, size, width, weight, and serifs [37], [46]. However, for the purpose of this research, we directed our focus towards two key font characteristics namely type and size. On the web, font type usually refers to the font family that represents a certain typeface, including all sizes and styles, for example, Arial and Verdana. To style a particular family type, the size of text may be increased or decreased using the font size attribute. There are different ways to resize the font including Ems (em), pixels (px), points (pt), and percent (%). In our font size analysis, we used pixels (px) as the sole measurement for the font since it is the main measurement unit for media shown on computer screens. Moreover, web designers usually use pixels as their key measurement unit for font size. Points, however, are used to measure font size for print media.

It is noted that more than half of the Arabic websites breached the principle of consistency when applying font types to the text. This means that multiple font types were used on the same web pages. Previous research, however, has already emphasized the role of consistency of user interface elements, such as font, in the overall usability of websites [35], [36]. On that basis, it is recommended that designers should adhere to the same font family to create attractive and consistent Arabic web designs.

Moreover, there was no evident consensus amongst the websites in favoring a particular font type in order to display the Arabic text. More than 35 font types were used by the selected 73 websites. There are various potential theories that could explain this staggering variability; firstly, the lack of research and thereby guidelines with respect to styling Arabic text to guide the effective design of Arabic websites; secondly, the weak conformance to international design practices and recommendations; thirdly, the lack of awareness that text formatting might ultimately impact text legibility and thereby readers' mental load and overall satisfaction.

Nonetheless, a pattern of font usage emerged within the Arabic websites; six font types recurred several times in the formatting of titles/headline, menus, and paragraphs. These font types were used differently according to the type of text element. For titles and menus, Droid Arabic Kufi, Tahoma, Helvetica, and Arial, emerged as favorites. It is no surprise that Droid Arabic Kufi came out as a winner for it is a unique font that reflects the Arabic calligraphy. This font gives a true feeling of the traditional Arabic scripts (Kufic scripts) that date back to the 7[th] century. For paragraphs, Tahoma, Helvetica, Arial and Times New Roman were used more frequently.

International recommendations suggest using font type Verdana since it is one of the most legible fonts as rated by online readers of English websites [38]. Nonetheless, this font did not appear frequently in the Arabic websites that we tested. Moreover, other famous fonts that were found earlier to be legible or attractive by English readers, such as Georgia [18] and Courier [10], did not rank well in the Arabic websites. However, Arial emerged as one of the top six font types used by the Arabic websites. Previous research confirmed that font type Arial was read significantly faster and was perceived as legible [10].

Design guidelines have indicated that font size 12 points (~ 16 pixels) is the most suited size for displaying text on English websites [39]. This is somewhat disputable as subsequent research suggested the use of font size 14 points (~19 pixels) instead. For displaying the Arabic text, recommendations are inconclusive yet. For example, [17] proposed to use font 14 points (~19 pixels) whereas [4] argued that font size 16 (22 pixels) to 18 points (~24 pixels) must be used for better readability. These suggestions are in strike contrast with the current practices of the Arabic web design. More than 60% of the inspected websites used font size ranging between 12 pixels to 16 pixels (i.e. 9 points to 12 points) to style the paragraphs. Similarly, menu items are sized from 13 pixels to 14 pixels, which is still considerably illegible [17].

In respect to using images in web design, we aimed at answering the following question. What types of images are incorporated within the Saudi websites given the Saudi culture? All Arabic websites included images as part of their content. Authors in [40] and [41] suggested that graphics improve the overall attractiveness of websites. Moreover, 90% of the Arabic websites we tested included human images. This is known to improve user engagement, trust, and satisfaction [21]. However, 80% of the Arabic websites used conservative images that respect the Islamic identity and cultural values of the Saudi people. Research studies agree that web designs should adhere to the cultural values of its users to achieve success and acceptance [42], [43], [44]. Two-thirds of the websites used images of both genders. The remaining third, however, used images of men only. This could be explained by the masculine dimension of the Saudi society as indicated by Hofstede scores [45]. More than 80% of the websites included images containing groups of people. Again, this can be justified by the collectivism dimension of the Saudi society [45]. Finally, there was high usage of animations, possibly, in an attempt to engage web visitors and improve the aesthetics of the websites. Evidently, the high presence of images in these websites contributed to lower performance scores and increased site loading time. Optimization of images' size, using the appropriate image compression techniques, is therefore recommended as a result of our analysis.

## VII. CONCLUSIONS

The 73 top viewed Arabic websites from local and international service providers, covering a wide range of areas in Saudi Arabia such as government, news, and e-commerce, were carefully inspected with a focus on the fonts applied to format the Arabic text and images incorporated to improve the overall site attractiveness. The Arabic websites suffered from inconsistency for they mixed font types on the same web pages. The added value of such practice is ambiguous and invites the designers of Arabic websites to pay more attention to the application of fonts more consistently within and across pages.

Droid Arabic Kufi was dominant in formatting the titles and menus of the Arabic sites. This font type reflects the specific characteristics of the Arabic writing, especially that related to calligraphy. Nevertheless, Tahoma and Helvetica were also used frequently to style the titles and menus. Surprisingly, however, the analysis showed that famous fonts used to style English text, such as Tahoma, Helvetica, Arial and Times New Roman, are used to format the Arabic paragraphs. Nonetheless, a myriad of other font types was used to style the Arabic text, which shows that the Arabic websites still do not favor a particular font type to improve text legibility or attractiveness. Instead, web designers of Arabic websites seem to trust what international websites frequently use such as Helvetica and Arial. This might not be necessarily efficient for reading the Arabic characters and language. Arabic content on the web is still yet to find its appropriate font type and font size.

Images were extensively used in the Arabic websites, which may indeed increase user engagement and visual appeal. However, the high number of images resulted in lower website loading speed, which is a key quality factor in web design. The use of images in the Arabic websites revealed three main characteristics, masculinity, collectivism and conservatism. These characteristics are in agreement with the Saudi culture and its religious identity.

Despite the insights highlighted by the results, there are several qualifying limitations that have to be enumerated here. The aforementioned studies describe the current design practices pertaining to font and image usage within the Arabic websites and do not shed light on the rationale for their use nor do they test real users' perception and judgment towards these practices. In other words, usability inspection studies do not allow obtaining a deep understanding of the effects of certain design decisions on user judgment and satisfaction. Follow-up studies are recommended to gauge user views and acceptability towards font and image usage in the Arabic websites. The inspection raised several interesting observations, but more statistical analysis is needed to assert differences between website categories.

## VIII. FUTURE RESEARCH DIRECTIONS

The herewith research provides a detailed outlook on the general practices of local and international websites in relation to the application of font and images in Arabic web design. To the best of our knowledge, this is the first study that investigates the use of font and images in Arabic websites extensively. Indeed, there is an overarching necessity to undertake more organized research focusing on these two important design elements. As part of a long-term agenda, we have already started to conduct a series controlled experiments in order to identify the most efficient Arabic font type and size to use in e-commerce websites as well as users' perception towards conservative and non-conservative images within fashion online stores and study how these images could influence customers' buying decisions. Researchers into the impact of typography on Arabic web usability should consider using eye-tracking to collect objective measures about Arabic reading performance and mental load. However, for the use of images, more user testing is required to judge users' attitudes and acceptance towards the shift in using less conservative images, especially in Arabic e-commerce websites, where buying decisions are greatly driven by the display of products such as clothes. Other interactive media such as animations and videos and their effect on user judgment and website performance would also need to be examined in the future.

REFERENCES

[1] E. Deborah, E-P. Rosen, "Website design: Viewing the web as a cognitive landscape", Journal of Business Research, Vol. 57, no. 7, pp. 787-794, 2004.

[2] G. Lindgaard, G. Fernandes, & C. Dudek, J. M. Brown, "Attention web designers: You have 50 milliseconds to make a good first impression!", Behaviour and Information Technology Journal, vol. 25 no. 2, pp. 115-126, 2006

[3] R. Sinha, M. Hearst, M. Ivory, and M. Draisin. "Content or graphics? an empirical analysis of criteria for award-winning websites", in Proc. Human Factors and the Web, 7th conference , Madison, WI, June 2001.

[4] A A Abubaker, J Lu ," The optimum font size and type for students aged 9-12 reading Arabic characters on screen: A case study", Journal of Physics: Conference Series, 2012, available at: https://iopscience. iop. org/article/10. 1088/1742-6596/364/1/012115/meta

[5] M. Bernard, "Criteria for optimal web design (designing for usability)", available at:http://psychology. wichita. edu/optimalweb/structure. htm, 8. 7. 2004.

[6] J. Nielsen, "Usability 101: Introduction to Usability, Available at:http://www. nngroup. com/articles/usability-101-introduction-to-usability/ ,2012.

[7] M. Y. Ivory, M. A. Hearst, "Improving website design",University of California, Berkeley, Vol. 01, Page no. 7, 2002.

[8] H. Holden, R. Rada, "Understanding the Influence of Perceived Usability and Technology Self-Efficacy on Teachers' Technology Acceptance", Journal Of Research On Technology In Education, (International Society For Technology In Education), vol. 43, no. 4, pp. 343-367, 2011.

[9] A. Seffah, M. Donyaee, K. R. B Kline, H. K. Padda, "Usability measurement and metrics: A consolidated model". Software Quality Journal. Vol. 14. pp. 159-178. Jun 2006.

[10] M. Bernard, B. Lida ,S. Riley, T. Hackler, K. Janzen, "A comparison of popular online fonts: which size and type is best?", Usability News. 2002 Jan;Vol. 4(No. 1) Online journal

[11] J. W. Palmer, "Web site usability, design, and performance metrics", Information Systems Research"; vol. 13, no. 2; pp. 15, Jun 2002.

[12] C. Flavian, R. Gurrea, C. Orus, "Web design: A key factor for the website success", J. Systems and IT. pp. 168-184. 2009.

[13] J. Hu, K. Shima, R. Oehlmann, J. Zhao, Y. Takemura, KI. Matsumoto, "An empirical study of audience impressions of B2C web pages in Japan, China and the UK", Electronic Commerce Research and Applications. , vol. 3, no. 2, pp 176-89 Jun. 2004.

[14] H. Shahizan, L. Feng, "Evaluating the Usability and Content Usefulness of Web Sites: A Benchmarking approach," Journal of Electronic Commerce in Organizations, vol. 3, no. 2, pg. 46, Apr-Jun 2005.

[15] D. Beymer, D. Russell,, P. Orton, " An eye tracking study of how font size and type influence online reading", Proceedings of the 22nd British HCI Group Annual Conference on HCI 2008: People and Computers XXII: Culture, Creativity, Interaction Vol. 2, pp. 15-18. Sept 2008.

[16] M. Gasser, J. Boeke,M. Haffernan,R. Tan, "The Influence of Font Type on Information Recall", North Am J Psychol, vol. 7, no. 2, pp. 181-188, 2005.

[17] A. Z. Alotaibi, "The effect of font size and type on reading performance with Arabic words in normally sighted and simulated cataract subjects", Clinical and Experimental Optometry, vol. 90, no. 3, pp. 203-206, 2007.

[18] M. A. Zamzuri, A, Mohamad, W. Rahani, S. Khairulanuar, Z. I. Muhammad, " Reading on the Computer Screen: Does Font Type has

Effects on Web Text Readability?", International Education Studies; Vol. 6, No. 3, pp. 26-35; 2013.

[19] N. Hojjati, B. Muniandy, "The Effects of Font Type and Spacing of Text for Online Readability and Performance ", Contemporary Educational Technology, vol. 5, no. 2,pp. 161-174, 2014

[20] A. Sears, J. A. Jacko, E. M. Dubach, "International Aspects of World Wide Web Usability and the Role of High-End Graphical Enhancements", Int. J. Hum. Comput. Interaction, vol. 12, no. 2, pp. 241-261, 2000.

[21] D. Cyr, H. Larios, M. Head, B. Pan, "Exploring human images in website design: a multi-method approach", MIS Quarterly Vol. 33 No. 3, September 2009.

[22] S. A. Becker and F. E. Mottay, "A global perspective on Web site usability," in IEEE Software, vol. 18, no. 1, pp. 54-61, Jan. -Feb. 2001.

[23] N. Bevan, L. Spinhof, "Are guidelines and standards for web usability comprehensive?", Proceedings HCI International 07, Springer, pp. 407-419, July 2007.

[24] N. Bevan, " Guidelines and Standards for Web Usability", Proceedings of HCI International, Lawrence Erlbaum, The Management of Information: E-Business, the Web, and Mobile Computing, vol. 2, July 2005

[25] Mohamed Benaida and Abdallah Namoun, "Technical and Perceived Usability Issues in Arabic Educational Websites" International Journal of Advanced Computer Science and Applications(IJACSA), vol. 9, no. 5, 2018.

[26] S. H. Mustafa, L. F. Al-Zoua'bi, "Usability of the Academic Websites of Jordan's Universities, An Evaluation Study", 9th International Arab Conference for Information Technology(ACIT'2008).

[27] A. Fernandez, E. Insfran,S. AbrahÃao, "Usability evaluation methods for the web: A systematic mapping study". Information & Software Technology. Vol 53, pp. 789-817, Aug 2011.

[28] T. Conte, J. Massolar, E. Mendes, GH. Travassos, "Web usability inspection technique based on design perspectives", IET software. Apr, vol3 no. 2, pp. 106-23. Apr. 2009.

[29] R. Otaiza, C. Rusu, S. Roncagliolo, "Evaluating the usability of transactional web sites", In 2010 Third International Conference on Advances in Computer-Human Interactions, pp. 32-37, Feb. 2010. K-S. Kang, B. Corbitt, "Effectiveness of Graphical Components in Web Site E-commerce Application-A Cultural Perspective", EJISDC, vol. 7, no. 2, pp. 1-6. 2001.

[30] "Top Sites in Saudi Arabia", available online at: https://www. alexa. com/topsites/countries/SA, last visit: 30. 3. 2019.

[31] "Fontface Ninja", available online at: https://fontface. ninja/, last visit: 30. 3. 2019.

[32] "Website Speed Test Image Analysis Tool", available online at: https://webspeedtest. cloudinary. com/, last visit: 30. 3. 2019.

[33] S. Mentes,, A. h. Turan, "Assessing the usability of university websites: An empirical study on Namik Kemal University". Turkish Online Journal of Educational Technology. Vol. 11, pp 61-69, Jul 2012.

[34] A. El-Sawy, M. Loey and H. El-Bakry, "Arabic Handwritten Characters Recognition Using Convolutional Neural Network,"WSEAS Transactions on Computer Research ,vol. 5, pp. 11-19, 2017.

[35] S. Steinau, O. Díaz, JJ. Rodríguez, F. Ibáñez, "A Tool for Assessing the Consistency of Websites", InICEIS (pp. 691-698). 2002.

[36] E. Shawgi, N. A. Noureldien, "Usability measurement model (umm): a new model for measuring websites usability", International Journal of Information Science, vol. 5. no. 1, pp. 5-13, 2015.

[37] T. Rabinowitz, "Exploring typography", Cengage Learning; 2015.

[38] J. Banerjee, D. Majumdar, D. Majumdar, "Readability, Subjective Preference and Mental Workload Studies on Young Indian Adults for Selection of Optimum Font Type and Size during Onscreen Reading", Al Ameen Journal of Medical Sciences. 4,pp. 131-143, Jan. 2011.

[39] M Bernard, M. Mills, "So, What Size and Type of Font Should I Use on My Website?", usability news, , Vol. 2, no. 2, July 2000.

[40] Y. -C. Lin, C. Yehb, C. Wei, " How will the use of graphics affect visual aesthetics? A user-centered approach for web page design , "et al. / Int. J. Human-Computer Studies vol. 71, pp. 217–227, 2013.

[41] D. Willis," Effects of using enhancing visual elements in Web site design", American Communication Journal,vol 3, no. 1, 1999, avalibale at: http://ac-journal. org/journal/vol3/Iss1/articles/Willis. htm.

[42] N. Tsikriktsis, " Does Culture Influence Web Site Quality Expectations? An Empirical Study", Journal of Service Research. Vol 5, no. 2, Nov 2002.

[43] R. Fletcher, "The impact of culture on web site content, design, and structure: An international and a multicultural perspective". Journal of Communication Management, vol. 10, no. 3, Jul. 2006.

[44] A. Marcus, E-W. Gould. "Crosscurrents: cultural dimensions and global Web user-interface design. " ACM Interactions, pp 32-46, vol. 7, no. 4, Jul. 2000.

[45] "Hofstede Insights", available online at: https://www.hofstede-insights.com/, last visit: 30. 3. 2019.

[46] J. Janbi, "Classifying Digital Arabic Fonts Based on Design Characteristics", 2018 21st Saudi Computer Society National Computer Conference (NCC), Riyadh, pp. 1-6. April 2018.

[47] Y. Huang, J. W. Wenjie Shi, "The impact of font choice on web pages: Relationship with willingness to pay and tourism motivation",Tourism Management, vol. 66, pp. 191-199, June 2018.

[48] M. A. Ababtain, A. R. Khan, "Towards a Framework for Usability of Arabic-English Websites", Procedia Computer Science, vol. 109, pp. 1010-1015, 2017.

[49] C. Yang, T. S. T. Huang, D. R. Shanks, "Perceptual fluency affects judgments of learning: The font size effect", Journal of Memory and Language, vol. 99, pp. 99-110, April 2018.

[50] J. Grobelny, R. Michalski, "The role of background color, interletter spacing, and font size on preferences in the digital presentation of a product", Computers in Human Behavior, vol. 43, pp.85-100, 2015.